\documentclass[BCOR=1mm, DIV=calc,10pt,
twoside=true,
headings=normal]{scrartcl}

\usepackage[utf8]{inputenc} 

\usepackage{url}

\usepackage{geometry} 
\usepackage{graphicx} 

\usepackage{booktabs} 
\usepackage{array} 
\usepackage{paralist} 
\usepackage{verbatim} 
\usepackage{subfig} 

\usepackage{fancyhdr} 
\usepackage{sectsty}
\allsectionsfont{\sffamily\mdseries\upshape} 

\usepackage[nottoc,notlof,notlot]{tocbibind} 
\usepackage[titles,subfigure]{tocloft} 

\title{ Enterprise AI Canvas \\ Integrating Artificial Intelligence into Business}
\author{Ulrich Kerzel \\
IUBH Internationale Hochschule GmbH \\
Juri-Gagarin-Ring 152 \\
D-99084 Erfurt }
\date{} 

\begin{document}

\maketitle

\begin{abstract}
Artificial Intelligence (AI)  and Machine Learning have enormous potential to transform businesses and disrupt entire industry sectors.
However, companies wishing to integrate algorithmic decisions into their face multiple challenges: They have to identify use-cases  in which 
artificial intelligence can create value, as well as decisions that can be supported or executed automatically. Furthermore, the organization 
will need to be transformed to be able to integrate AI based systems into their human work-force. Furthermore, the more technical aspects of 
the underlying machine learning model have to be discussed in terms of how they impact the various units of a business: Where do the relevant data come
from, which constraints have to be considered, how is the quality of the data and the prediction evaluated?

The Enterprise AI canvas is designed to bring Data Scientist and business expert together to discuss and define all relevant aspects which need
to be clarified in order to integrate AI based systems into a digital enterprise. It consists of two parts where part one focuses on the business view
and organizational aspects, whereas part two focuses on the underlying machine learning model and the data it uses.
\end{abstract}

{\textbf{keywords}: artificial intelligence, business, decision making, canvas, value proposition}

\section{Introduction}

Artificial Intelligence (AI)  and Machine Learning have enormous potential to transform businesses and disrupt entire industry sectors.
When referring to artificial intelligence, the distinction between general (or strong)  and narrow AI is generally made:  General AI systems \cite{Turing, MCCARTHY20071174} 
exhibit intelligent behavior en par or even surpassing human intelligence in a wide range of scenarios. While general AI systems are still elusive it is expected that they would interact and reason
much like we humans do. Narrow AI systems on the other hand aim to solve a particular problem and specialize in the execution of a specific, singular task. 
This approach has seen spectacular successes such as  playing the game ``Go" without human knowledge \cite{Silver2017} or cancer detection \cite{Esteva2017, Ardila2019}.

While the performance of narrow artificial intelligence is indeed astonishing in these and other examples, they mainly focus on a specific research questions.
Businesses on the other hand first and foremost need to address the question how to generate value for their customers and stake-holders. 
P. Sondergaard, Senior VP at Gartner said already in 2015 that``algorithms are where the real value lies [because] algorithms define action" \cite{Sondergaard}. 
This statement captures the essence of algorithmic decisions quite well: Algorithms allow to create value and can be tied to decisions: The output of an algorithm is 
typically a prediction, either a classification or a regression,  which can then be translated into operational decisions. While many companies also engage in fundamental research, 
the driving factor for most innovations and offerings can be traced back to the question: How does the company create value?

Integrating artificial intelligence into a business requires more consideration than traditional analytics: While integrating data-driven decisions is not new for many companies, 
these typically focus on establishing or expanding teams of specialists with sophisticated tools. As these activities are still based on human teams, they can be integrated
into organizations in familiar ways including management and organizational structures. However, integrating decisions taken by artificial intelligence also needs to address
the questions how human and  artificial intelligence work together and how  and by whom decisions are taken. From an organizational perspective, AI systems are not 
"managed". Furthermore, for the foreseeable future AI systems are not suitable for all kind of decisions: Some decisions are better taken by humans, others are more
suitable to be delegated to an artificial intelligence. Deciding which category a particular business idea falls into requires extensive technical expertise beyond the level
of most business expert. 

The Enterprise AI canvas is designed to bring Data Scientist, AI specialists and business experts together and aims to ask all relevant questions which need to 
be addressed to first identify the most promising use-cases which bring value to the company and its customers, as well as more technical detail regarding the underlying data and machine learning model.

\section{Relation to other work}
Many approaches exist to identify new business opportunities. A very successful approach is the Business Model Canvas (BMC), 
proposed by Osterwalder and Pigneur \cite{OsterwalderThesis, BusinessModelCanvas}. The Business Model Canvas used nine building blocks
which focus on the value proposition, key partners, activities, resources, as well as customers and cost and revenue streams.
The BMC is a helpful tools for business experts to identify new value propositions and how they bring value to the company and customers, 
however, as a general canvas, the BMC does not address the additional requirements of evaluating whether the proposition is suitable
for an AI based system or  integrating artificial intelligence into the organization. Implicitly, the BMC assumes that some team will be responsible of
realizing the value proposition - but the details do not need to be evaluated at this stage.
The Machine Learning Canvas (MLC) proposed by Dorard \cite{Dorard} on the other hand focuses mainly on the technical aspects of developing a 
use-case suitable for machine learning algorithms, covering model related aspects such as data sources, data collection, features as well as making
predictions and the resulting decisions or actions. While the MLC also contains the  ``value proposition" as central element, most of of the blocks of the canvas
are related to technical details which are more relevant to Data Scientists than to business experts, which makes the machine learning canvas mainly relevant
for the Data Science team developing and implementing a business idea.
The AI Canvas proposed by Agrawal {\em et al.} \cite{AICanvas_Agrawal} uses AI as a general term for machine based predictions, synonymly with machine learning. 
This canvas captures the main ideas of a proposed new project such as what should be predicted and the resulting action, as well as some general aspects
of required data and training but doesn't go into neither the technical details like the machine learning canvas nor how
businesses would integrate this in their operations.
The AI Canvas proposed by K. Dewalt \cite{AICanvas_Dewalt} provides a high-level mix of business and data science centric aspects. The business centric
aspects include discussions about {\em opportunity} ,  {\em solution}  and  {\em customers} , similar to {\em value proposition} in the machine learning canvas by L. Dorard.
Similar to the Machine Learning Canvas, this canvas also includes fields for {\em data sources}, {\em model development} and {\em success criteria}, though 
they remain on a higher level compared to the machine learning canvas.
Finally, the AI Project Canvas suggested by J. Zawadski \cite{AICanvas_Zawadszki} is based on the original business model canvas and focuses mainly
on the development of the business idea such as the {\em value proposition} cost and revenue, as well as who are the end customers and stakeholders and required skills.
However, the AI Project Canvas doesn't go in more technical details and is hence mainly aimed at project managers.

\section{The Enterprise AI Canvas}
At first glance it might be tempting to hand the Business Model Canvas  \cite{OsterwalderThesis, BusinessModelCanvas} or
the AI Project Canvas  \cite{AICanvas_Zawadszki} to business experts or project managers and then the Machine Learning Canvas  \cite{Dorard} to 
Data Scientists in the next step. This is indeed what happens in industry and application quite often, although the various canvases
are seldom used in practice and instead potential projects are discussed and derived in presentations or flip-charts.
However, in many practical situations this will not lead to satisfactory results: Developing data-driven applications using machine learning 
or (narrow) artificial intelligence requires a detailed understanding of available or obtainable data for a given use-case as well as an estimation
of the feasibility developing a model in reasonable time. While these aspects can be well understood by (senior) data scientists, business experts 
or project managers are often too far detached from these details. On the other hand, Data Scientists have typically only limited insight into 
how to create business value and optimize operational decisions in a corporate setting, as well as which organizational change or change management would 
be required to integrate an AI project in a company. Project ideas and details, technical aspects and organizational implications are intertwined and closely related.
Discussing the business aspects first and getting them approved by the senior management before engaging deeply with the data scientist can either lead to 
unnecessary loops re-defining projects at best and implementing something which fulfills the project plan but doesn't meet the requirements of the customer at worst.
In addition, business experts and data scientists have a very different background and use a different language or jargon.

The Enterprise AI canvas proposed in this work  is specifically designed such that all relevant communities 
such as business and domain experts as well as data scientists can work together on a common framework. This allows to capture the expertise
from a diverse project planning team and ensures that all relevant aspects of the whole project are considered. Special focus has been given
to the consideration of the decision making, decision optimization and impact on organizational structures. The technical details such as defining the prediction target
are  often discussed among data scientists only, these specifications have often far-reaching consequences on the business side: 
As operational decisions are taken by machines, how should the company deal with the people who 
made these decisions earlier? What will these people work on and how will their job role change?
A further important aspect is to consider how operational decisions are optimized. 
Prior to the AI based project, deriving  the optimal decision has typically been 
the responsibility of a manager or expert, as decisions are taken automatically, how is the optimization process changed to incorporate the predictions
from the AI?

\begin{figure}[h]
\begin{center}
\includegraphics[scale=0.5]{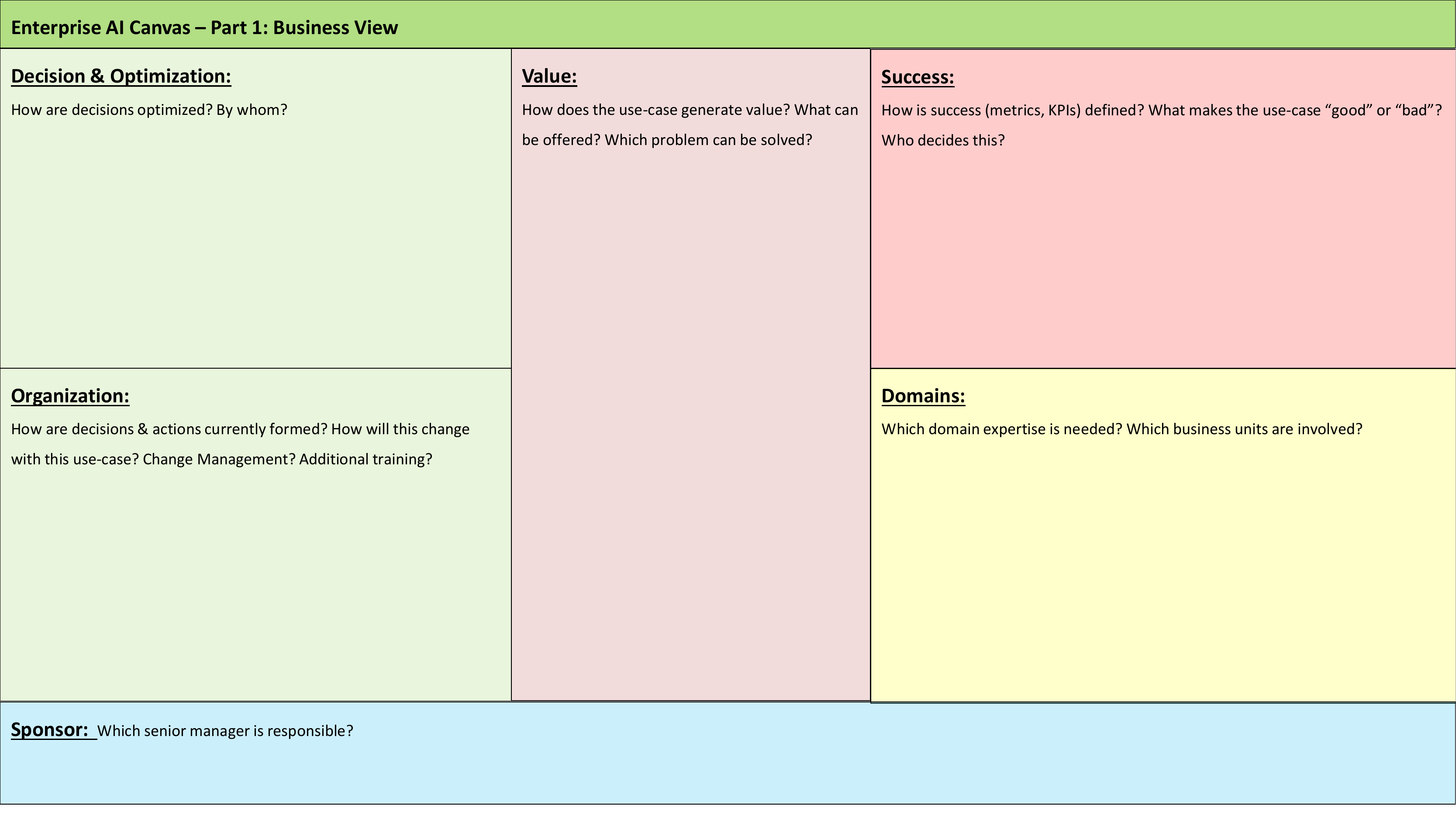}\\
\includegraphics[scale=0.5]{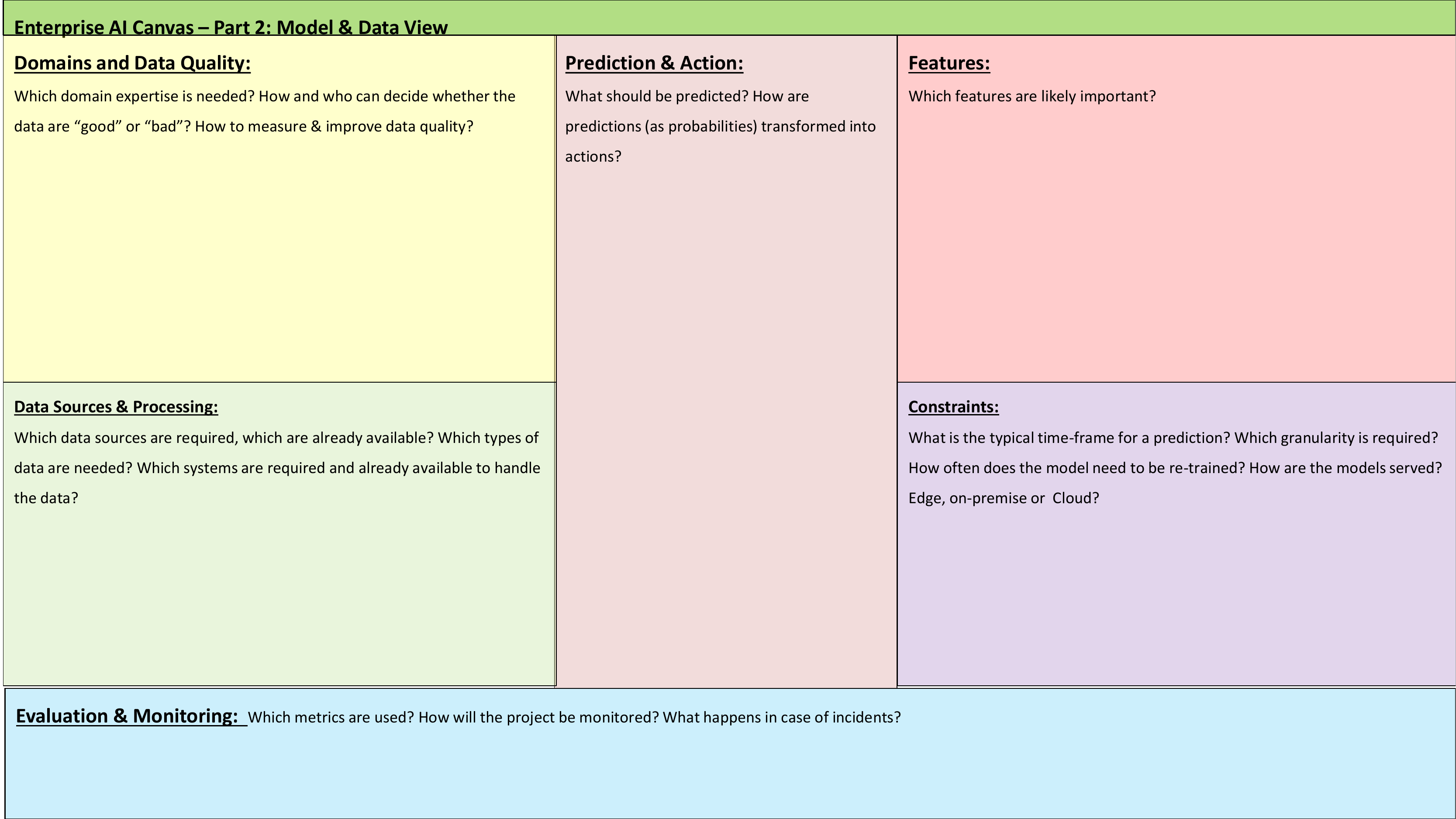}
\caption{\label{fig:AIcanvas} Enterprise AI Canvas}
\end{center}
\end{figure}

The Enterprise AI Canvas consists of two parts, the first part is mainly aimed at the business perspective of integrating an AI system, the 
second part focuses more on the underlying machine learning and data model. Both parts are shown in detail in fig.\ref{fig:AIcanvas}
and all elements are discussed below in detail.
Each element of the canvas can be considered separately in any order, however, it is advisable to start with the central elements first and then 
move towards the most closely related elements. Hence it is helpful to start with {\em Value} in part one and first consider how the potential new business opportunity 
can create value. Consequently, it is helpful to start with {\em Prediction \& Action} in part two to discuss what needs to be predicted exactly and how a prediction
by a machine learning model can be turned into an operational action.

\subsection{Part 1}
The first part of the Enterprise AI canvas focuses on the business view of the opportunity: How does the opportunity bring value, what is meant by "success", how does it impact the organization to deliver the opportunity?

\begin{itemize}
\item{\textbf{Value:} The central item of the potential business opportunity is how it creates value for the organization and the customers. The main relevant questions are focused on the potential offering and which problem can be solved with the new business opportunity. The potential use-case may be a new offering or a substantial improvement on previous approaches.}
\item{\textbf{Success:} The central part of a new business opportunity or a significant improvement of an existing offering is connected to its success. However, what {\em success} means needs to be carefully defined. From a technical perspective, Data Scientists may be tempted to focus on metrics which focus on the quality of the individual predictions made by the machine learning model. However, as these predictions are embedded into operational decisions, a "good" prediction does not necessarily translate into a "good" operational decision. Business and Data Science experts need to work together and identify specific metrics which allow to evaluate the quality of the operational decisions made by the AI model. These are likely not technical metrics
but Key Performance Indicators (KPIs) related to the business opportunity. Note that this does not imply that the technical metrics should not be used, however, there are often many layers between the prediction of a machine learning model and the resulting operational decision which need to be taken into account. What matters in the end is that the best possible decision is taken.}
\item{\textbf{Decisions \& Optimization:} This aspect focuses on the operational aspect of the decisions related to the potential use-case. If an existing approach is improved substantially using an AI based system, business specialists need to understand which decisions are taken at the moment and how are they formed. Which reasoning goes into making decisions? Ideally, decisions are already based on a clear approach or data-driven analytics, although many decisions may still be taken with "gut feeling". If a new business opportunity is explored, the associated decisions need to be made explicit and the objectives according to which the decisions are optimized need to be specified.}
\item{\textbf{Organization:} This item is related to the point {\em Decision \& Optimization} which focuses on which decisions need to be made in the context of a new business opportunity and how these decisions are optimized to achieve the best possible outcome. The item {\em Organization} focuses on who currently makes the decision: How are they formed? Who does that? What will happen once the AI based system is integrated in the process? Which decisions will be taken by humans, which by AI? What happens with the humans currently involved? Will they need additional training? Does part of the organization of event the whole organization need to be changed to accommodate the fact that operational decisions are now at least party taken
by a machine? Who will prepare the organization for this change and how? Failing to address these points can create significant anxieties among the employees such as the fear of loss of their jobs or similar. }
\item{\textbf{Sponsor:} Any new substantial change in the operations and organization of company requires careful consideration by the senior management team. This is particularly true in the case of system based on artificial intelligence which may likely have a large impact on the way the business works when human and artificial intelligence work together. It is therefore paramount to identify the senior management team which is responsible for this project to guide the company through the necessary change process.}
\item{\textbf{Domains:} In order to generate value, a new business opportunity has to be embedded in a specific industry domain or application area. When evaluating the prospects of a new opportunity it is important to integrate the relevant domain expertise from an early point on. This can relate both to internal expertise in various business units as well as external expertise.}
\end{itemize}

\subsection {Part 2}
Part 2 of the Enterprise AI canvas focuses more on the technical aspects of the implementation of the machine learning model as the foundation of the artificial intelligence based decision system.

\begin{itemize}
\item{\textbf{Prediction \& Action:} This item relates to{\em Decision \& Optimization} and {\em Value} in part one. Whereas the aspects in part one were mainly concerned with the business opportunity and how decisions are derived and implemented, the part {\em Prediction \& Action} in part two focuses on the concrete prediction the AI system needs to make to fulfill the objective, i.e. which quantity exactly needs to be predicted? Is this a classification problem or should a quantity be predicted (regression)? Fundamentally, all predictions are probabilities (in case of classifications) or probability density distributions (in case of regression). How are these probabilities transformed into actions? If e.g. a threshold needs to be defined, how is this value determined? }
\item{\textbf{Features:} Once it has been defined what exactly the model needs to predict, extensive consideration should be given to potential feature variables which are likely needed by the machine learning model to calculate the individual predictions. Although not every aspect can be considered at this stage as more feature variables might be deemed important during the development phase, this step helps to understand which data sources will be required. }
\item{\textbf{Data Sources \& Processing:} This part relates to {\em Features}: Once the major feature variables have been identified, the next step is to investigate which data sources can be accessed to extract the relevant data. Potential data sources may cover a wide range of internal and external sources. Ideally, the list should be ranked according to a first estimate of
the priority of access of these system as extracting data from each system takes time and effort and the most important data sources should be handled first.
Furthermore, the AI system itself also requires adequate processing capabilities both for handling the data and for the calculation of the predictions. The required computing and storage capacity need to be considered as part of the project planning, as well as where these systems are operated, such as a private data-center or a cloud-based system.}
\item{\textbf{Domains \& Data Quality}: Ultimately, an AI based system is only as good as the data it can use to learn from. 
 Consequently, the quality of the data is paramount and significant effort should be spent to clean the data before it is used to train an AI system. Although often mentioned in popular belief, data don't speak for themselves but deep domain expertise is required to decide whether a particular data point is correct or whether an outlier is likely due to a technical issue or is related to a rare but valid event. Which expertise is needed to assess data quality? Which criteria can be defined to measure and quantify the quality of the data and which processes can be established to improve the quality of the data throughout the project ? }
\item{\textbf{Constraints:} In most practical applications a number of constraints have to be met. For example, the predictions have to be calculated within a specific time or particular requirements have to be met regarding data security and data privacy. In some cases this may imply that cloud based services are the best approach, in other cloud based services may be ruled out and dedicated data centers or edge devices have to be used.}
\item{\textbf{Evaluation \& Monitoring:} This part is related to {\em Success} from part one of the canvas. After defining how the project is evaluated it now needs to be discussed how the relevant metrics will be calculated and presented. Ideally, a first plan is made how to handle any incidents or deviations from the allowed range of metrics.}
\end{itemize}

\section{Example}

In the following example, a fictitious supermarket chain was analyzed and the Enterprise AI canvas filled in accordingly.  The example was chosen because the different elements are easily relatable to everyday experiences and because of the author's experience in this specific area. In general, supermarkets are offering a wide range of products to end customers where the range of products (assortments) varies from hundreds or thousands of different products for discounters to tens of thousand different products for general supermarkets. Typically, supermarkets get new goods once per day or every few days. Many products have a long shelf-life, such as  frozen, canned or dry food, and hence replenishing these products is generally not critical as irregularities in the shipment can be compensated in the next cycle as long as sufficient stock is available on the shelf. However, a wide range of products has only a short shelf-life and needs to be disposed of once the products perish. These include for example fresh produce, flowers, bakery products, dairy, meat and similar. Some of these products (e.g. fresh bread) can only be sold during a single day. These products, called fresh or ultra-fresh, are much more challenging  from a replenishment perspective: They can't be stocked to compensate for an unexpected surge in demand or a missed shipment and keeping a large safety stock leads to high level of waste disposal which is both costly and harmful for the environment. Keeping the stock too low on the other hand will result in stock-out situations where customers face empty shelves. 
Therefore, supermarkets increasingly turn to automatic ordering systems based on artificial intelligence which predict the expected demand of individual products and optimize the shipments to each store \cite{BYMorrison}. Such a system has the benefit that it can take all available information into account from past sales records to weather information, as well as constraints from lot sizes and shipping schedules. 

\begin{itemize}
\item{\textbf{Value:} Optimized replenishment decisions are beneficial for both the customer and the retailer: The customer face fewer stock-out situations in which they do not find the desired goods. In addition, the retailer can reduce the amount of waste due to perished products and reduce the overall inventory.}
\item{\textbf{Success:} The success of the replenishment decisions can be measured by operational key performance indicators (KPIs), in particular the waste rate, i.e. the amount of perished products which have to be disposed of and the stock-out rate, i.e. the number of times products are not available on the shelves for customers to buy. In principle, the overall inventory is a 
third metric to optimize, however, in times of low interest rates the latter optimization is often less of a concern as long as the excess inventory does not perish quickly and fits into existing storage capacities. These metrics need to be decided upon both by the business and data science experts: Business experts need to determine which Kips are relevant for the project
and together with the data scientists they need to identify which range of values are possible due to the constraints from e.g. lot-size or shipping schedule and which range of KPIs 
is considered to be successful for the project.}
\item{\textbf{Decisions \& Optimization:} In this example, the operational decisions taken are the order quantity per product per delivery cycle and store. Following the implementation of the project, all decisions should be taken by the AI based system with human oversight to react to unforeseen circumstances such as extreme weather, food scandals, etc.}
\item{\textbf{Organization:} This part of the canvas has been intentionally left blank as this varies significantly between different retailers. However, typically retailers move from a manual
system supported by spreadsheets where each decision was taken by a human operator to an automatic system. This generally requires both change management efforts to prepare the company as a whole to this shift, as well as training of the individual operators whose role is changing due to the implementation of the project.}
\item{\textbf{Sponsor:} As the change of the replenishment process changes, core business processes are affected which should be generally sponsored and monitored by the CEO. However, a member of the senior management team should be assigned in addition to focus on the operational responsibilities of the project.}
\item{\textbf{Domains:} Successful implementation of such a project will require the cooperation and interaction with several departments such as product management, marketing and supply chain experts. Since several critical departments need to work together, the sponsor needs to be sufficiently senior to bring all required departments together.}
\item{\textbf{Prediction \& Action:}The starting point of the replenishment is the prediction of the future demand per product per order cycle per store. This quantity is ideally 
predicted as a full probability density distribution from which the optimal point estimator is derived which forms the basis for the later ordering decision. Typically, these 
predictions are needed one day ahead for operational purposes and 2-30 days ahead or longer to anticipate future trends and as a fail-safe precaution: Should the prediction not be
available for any reason, these longer term prediction can be used as a basis for the subsequent ordering decision instead. In order to derive the optimal point estimator, 
a simulation can be used in which each quantile of the predicted probability density distribution forms the basis of a simulated order and the effect on the metrics chosen
in the element {\em Success} can be studied. }
\item{\textbf{Features:} Previous sales records will be the most important information for future demand, assuming the underlying consumer behavior hasn't changed drastically.
In addition, properties of the individual products influence buying decisions as well as the location, size and other store characteristics due their effect on customer segmentation. Current information like weather forecasts or special calendar events such as holiday, festive seasons or major sports events will influence demand as well. }
\item{\textbf{Data Sources \& Processing:} Historic sales data will be in some form of a data warehouse or database, new records are either added after the store closes and the data are pulled from the till or the tills stream each completed transaction into the relevant network. Marketing information, product data, etc will be in some Enterprise Resource Planning (ERP) tool. In addition, external information has to be brokered from information providers such as weather forecast, calendar information, etc. All data have to be made available for the training phase. }
\item{\textbf{Evaluation \& Monitoring:} Customer and business centric metrics as defined under {\em Success} will be monitored at least on a daily basis, more frequently if the data are available. In addition, further metrics such as the mean absolute deviation (MAD) can be used to evaluate the quality of the prediction. A profile histogram showing the behavior of the predicted point estimator against the observed sales can help to determine if there are any biases.} 
\end{itemize}

\begin{figure}[h]
\begin{center}
\includegraphics[scale=0.5]{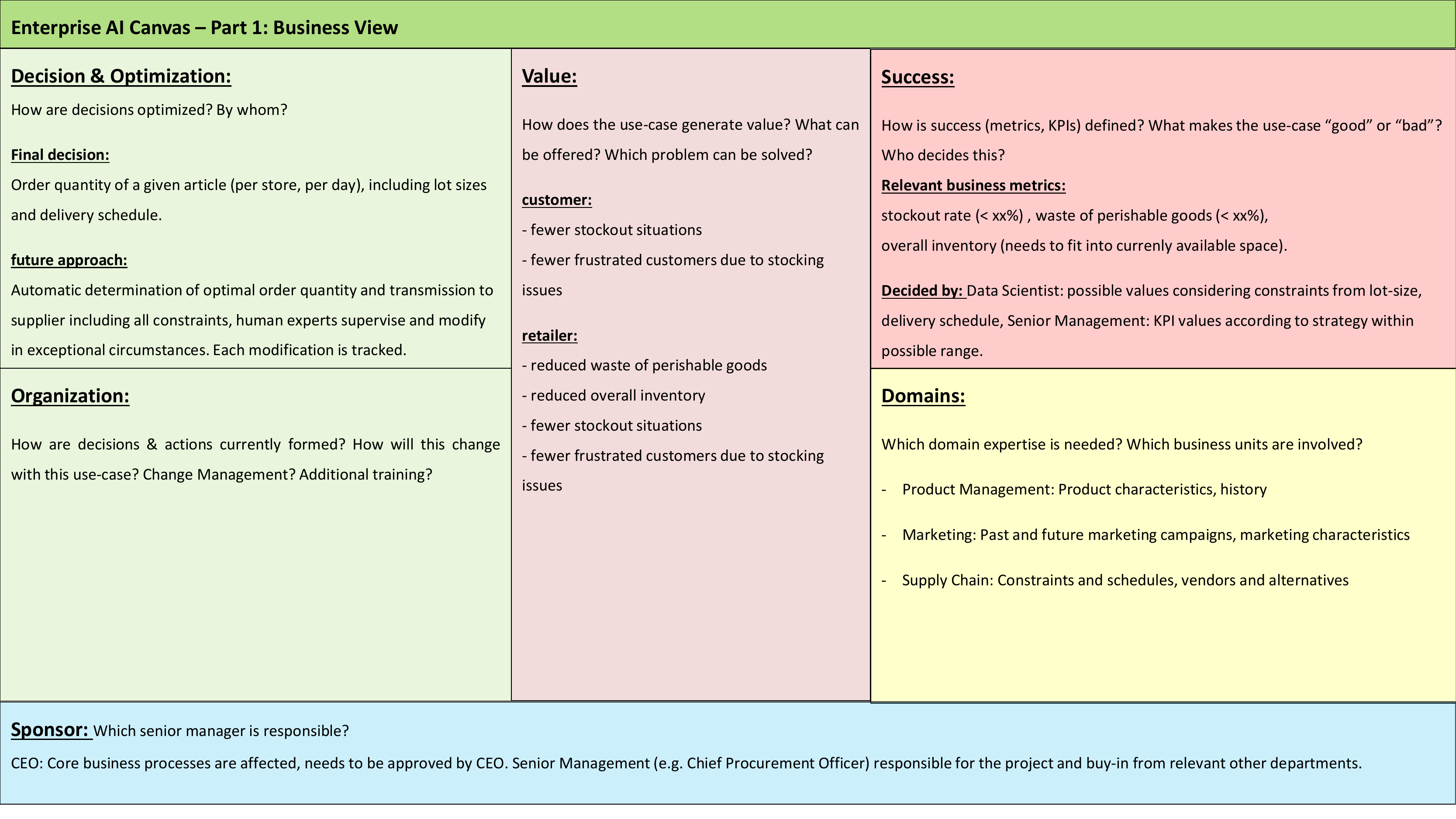}\\
\includegraphics[scale=0.5]{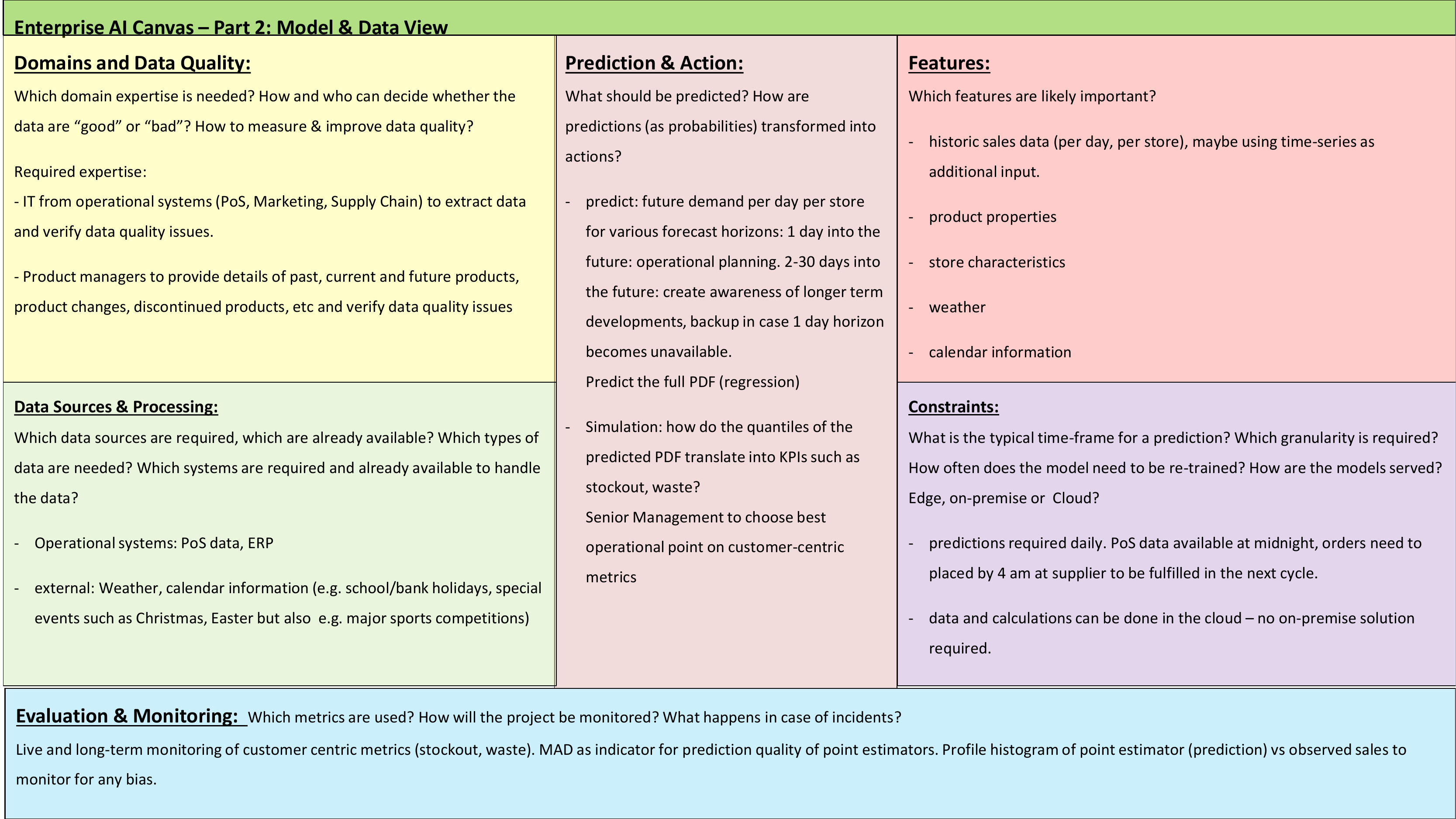}
\caption{\label{fig:AIcanvasExample} Example: Enterprise AI Canvas for a supermarket chain focusing on store replenishment.}
\end{center}
\end{figure}

\section{Conclusion}
Recent advances in artificial intelligence are set to disrupt a wide range of industries and sectors. However, companies wishing to integrate AI based systems 
into their offering or operations face significant challenges how to proceed. The Enterprise AI canvas allows to bring business and data science
experts together and systematically evaluate potentially new business opportunities. The canvas consists of two parts, where part one
focuses on the value a new AI based offering or operation would bring to the business whereas part two focuses more on the technical 
aspects of the underlying machine learning model. Both parts are intertwined and relate to each other to allow an optimal 
evaluation of business and data science related aspects of a new project.

\bibliography{EnterpriseAICanvas}

\begin{thebibliography}{10}

\bibitem{Turing}
A.~M. Turing, ``Lecture to the london mathematical society,'' {\em The
  Collected Works of A.M. Turing}, vol.~Mechanical Intelligence, 1947.

\bibitem{MCCARTHY20071174}
J.~McCarthy, ``From here to human-level ai,'' {\em Artificial Intelligence},
  vol.~171, no.~18, pp.~1174 -- 1182, 2007.
\newblock Special Review Issue.

\bibitem{Silver2017}
D.~Silver, J.~Schrittwieser, K.~Simonyan, I.~Antonoglou, A.~Huang, A.~Guez,
  T.~Hubert, L.~Baker, M.~Lai, A.~Bolton, Y.~Chen, T.~Lillicrap, F.~Hui,
  L.~Sifre, G.~van~den Driessche, T.~Graepel, and D.~Hassabis, ``Mastering the
  game of go without human knowledge,'' {\em Nature}, vol.~550, pp.~354 EP --,
  Oct 2017.
\newblock Article.

\bibitem{Esteva2017}
A.~Esteva, B.~Kuprel, R.~A. Novoa, J.~Ko, S.~M. Swetter, H.~M. Blau, and
  S.~Thrun, ``Dermatologist-level classification of skin cancer with deep
  neural networks,'' {\em Nature}, vol.~542, pp.~115 EP --, Jan 2017.

\bibitem{Ardila2019}
D.~Ardila, A.~P. Kiraly, S.~Bharadwaj, B.~Choi, J.~J. Reicher, L.~Peng, D.~Tse,
  M.~Etemadi, W.~Ye, G.~Corrado, D.~P. Naidich, and S.~Shetty, ``End-to-end
  lung cancer screening with three-dimensional deep learning on low-dose chest
  computed tomography,'' {\em Nature Medicine}, 2019.

\bibitem{Sondergaard}
H.~Levy, ``The arrival of algorithmic business.''
  \url{https://www.gartner.com/smarterwithgartner/the-arrival-of-algorithmic-business/}.
\newblock Last accessed on 11. Sep. 2020.

\bibitem{OsterwalderThesis}
A.~Osterwalder, {\em The Business Model Ontologya Proposition in a Design
  Science Approach}.
\newblock PhD thesis, University of Lausanne, CH, 2004.

\bibitem{BusinessModelCanvas}
A.~Osterawalder and Y.~Pigneur, {\em Business Model Generation}.
\newblock Wiley, 2010.

\bibitem{Dorard}
L.~Dorard, ``The machine learning canvas.''
  \url{https://www.louisdorard.com/machine-learning-canvas}.
\newblock Last accessed on 11. Sep. 2020.

\bibitem{AICanvas_Agrawal}
A.~Agrawal, J.~Gans, and A.~Goldfarb, ``A simple tool to start making decisions
  with the help of ai.''
  \url{https://hbr.org/2018/04/a-simple-tool-to-start-making-decisions-with-the-help-of-ai}.
\newblock Last accessed on 11. Sep. 2020.

\bibitem{AICanvas_Dewalt}
K.~Dewalt, ``Become an ai company in 90 days.'' ISBN 978-0-692-19233-7.

\bibitem{AICanvas_Zawadszki}
J.~Zawadzki, ``Introducing the ai project canvas.''
  \url{https://towardsdatascience.com/introducing-the-ai-project-canvas-e88e29eb7024}.
\newblock Last accessed on 11. Sep. 2020.

\bibitem{BYMorrison}
E.~Robinson.
  \url{https://www.technologyrecord.com/Article/morrisons-implements-blue-yonders-ai-stock-replenishment-technology-62728
  }.
\newblock Last accessed on 11. Sep. 2020.

\end{thebibliography}
\bibliographystyle{ieeetr}

\end{document}